\documentclass[aip, jmp, amsmath,amssymb, reprint,floatfix,]{revtex4-1}

\usepackage{import}
\usepackage{graphicx}
\graphicspath{ {./figures/} }
\usepackage{mathtools}
\usepackage{dcolumn}
\usepackage{bm}
\usepackage{braket}
\usepackage{enumitem}
\usepackage{float}
\usepackage[section]{placeins} 
\usepackage{booktabs}
\usepackage{array}
\begin{document}



\title{Alchemical geometry relaxation}

\author{Giorgio Domenichini} 
\affiliation{Faculty of Physics,
University of Vienna,
Kolingasse 14-16,
1090 Vienna,
Austria}

\author{O.Anatole von Lilienfeld}
\email{anatole.vonlilienfeld@univie.ac.at}
\affiliation{Faculty of Physics,
University of Vienna,
Kolingasse 14-16, 1090 Vienna, Austria}
\affiliation{Institute of Physical Chemistry and National Center for Computational Design and Discovery of Novel Materials (MARVEL), Department of Chemistry, University of Basel, Klingelbergstrasse 80, 4056 Basel, Switzerland}
\begin{abstract}
We propose to relax geometries throughout chemical compound space (CCS) using alchemical perturbation density functional theory (APDFT). 
APDFT refers to perturbation theory involving changes in nuclear charges within approximate solutions to Schr\"odinger's equation. 
We give an analytical formula to calculate the mixed second order energy derivatives with respect to both, nuclear charges and nuclear positions (named "alchemical force"), within the restricted Hartree-Fock case.
We have implemented and studied the formula for its use in geometry relaxation of various reference and  target molecules.
We have also analysed the convergence of the alchemical force perturbation series, as well as basis set effects.
Interpolating alchemically predicted energies, forces, and Hessian to a Morse potential yields more accurate geometries and equilibrium energies than when performing a standard Newton Raphson step. 
Our numerical predictions for small molecules including BF, CO, N2, CH$_4$, NH$_3$, H$_2$O, and HF yield mean absolute errors of of equilibrium energies and bond lengths smaller than 10 mHa and 0.01 Bohr for 4$^\text{th}$ order APDFT predictions, respectively.
Our alchemical geometry relaxation still preserves the combinatorial efficiency of APDFT: Based on a single coupled perturbed Hartree Fock derivative for benzene we provide numerical predictions of equilibrium energies and relaxed structures of all the 17 iso-electronic charge-netural BN-doped mutants with averaged absolute deviations of $\sim$27 mHa and $\sim$0.12 Bohr, respectively. 
\end{abstract}
\maketitle
 \section{Introduction}
Chemical compound space is conceptually defined as the infinite set of all possible chemical compounds \cite{ccs,ccs2}.
Extensive knowledge of chemical compound space based on quantum mechanical calculations only is made extremely challenging by the unfathomably large number of molecules that have to be taken into consideration; in practical terms, only a small portion of it can be screened using standard computational methods.
In order to improve our understanding of chemical compound space, as well as to enhance molecular design efforts, faster methods to scan chemical space are needed \cite{ccsLilienfeld,freeze2019search_Batista}.
This can be achieved in two possible ways: either using an exact method at a low level of theory (cheap QM methods or force fields) or using an approximate method for estimating high-level theory calculations.
Both quantum machine learning (QML) and alchemical perturbation density functional theory (APDFT) take advantage of data previously acquired, and generate new predictions using either interpolation (QML) or extrapolation (APDFT) techniques.
QML  \cite{qml_ccs_anatole2018, qml_nutshell_rupp2015} can be used effectively to predict several quantum properties \cite{von2020exploring,qml_rupp_atomization, faber_qml_lower_DFT,christensen2019operators,weinreich2021machine,qml_properties}, recent works \cite{qml_optimization_hammer, qml_lemm2021energy, Keith2021QML} succeeded in machine learning the optimized molecular geometries, but most QML methods do not predict directly geometrical structures.

APDFT \cite{apdft} is an alternative to QML for the rapid screening of chemical space. 
QML comes at the initial cost of securing a large training set, and this is not always available for the desired level of theory, the desired property of interest, or a relevant chemical subspace; in this regard, APDFT is more versatile because it requires the explicit QM calculation of only one reference molecule. 
From the reference molecule calculation and through the use of alchemical perturbations, it is possible to subsequently obtain estimates of molecular properties for a combinatorially large number of target molecules \cite{apdft}.
APDFT accuracy depends on the perturbation order and on the basis set choice \cite{Domenichini2020}. If a high-quality calculation (e.g. CCSD) is used as reference, the error in third-order energy prediction can be smaller than the error made using a cheaper QM method such as HF and MP2. \cite{Domenichini2020,apdft}


Alchemical changes and derivatives were at first introduced as a way to rationalize energy differences between molecules.\cite{4DEDF,politzer1974some,Levy1978}
Successive works showed that APDFT is also capable of predicting many other molecular properties, such as electron densities, dipole moments \cite{apdft}, ionization potentials \cite{von2006molecular, marcon2007tuning}, HOMO energy eigenvalues \cite{Lilienfeld2009}, band structures\cite{Chang2018}, deprotonation energies\cite{Rudorff2020,Cardenas2020Deprotonation}.
APDFT is particularly convenient if applied to highly symmetric systems where the explicit calculation of only a few derivatives (all  the others can be obtained by symmetry operations) is required, e.g., in the BN-doping of polycyclic aromatic hydrocarbons.\cite{benzene2013Geerlings, Lilienfelds_bn_doped_graphene,c60_2018Geerlings}
Using pseudopotentials, quantum alchemy can also be applied to crystal structures \cite{marzari_94, saitta1998structural,bellaiche2002low, Lilienfeld2010, Solovyeva_crystals_2016, Chang2018}, and can even suggest catalyst improvements \cite{Griego2020MLcAP_Keith,Keith2017,Keith2019,Keith2020, Griego_Keith2021_comp_guidelines}. 


One of the major remaining challenges in APDFT is to include geometry relaxation as the equilibrium molecular structure changes from reference to target; this would constitute a huge improvement over  fixed atoms "vertical" predictions. 
To the best of our knowledge, only two papers \cite{Chang_bonds, Lilienfeld2009} have tried to predict equilibrium energies, applying APDFT to "non-vertical" energy predictions (predicting energies for a perturbation that includes both changes in nuclear charges and in geometry). 

\begin{figure}[h]
    \centering
    \includegraphics[width=\linewidth]{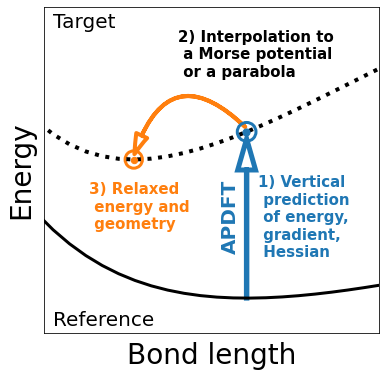}
    \caption{APDFT predictions of gradients and Hessians enable efficient geometry relaxation of target molecules. The predictive accuracy improves with Morse potential interpolation.
    }
    \label{fig:Morse_diatomics}
\end{figure}

In this paper we propose an alternative approach: we predict the geometry energy derivatives (gradients and Hessians) for the target molecule, and subsequently, use them to relax the geometry (see Fig.\ref{fig:Morse_diatomics}).

Alchemical forces are defined as the second-order mixed derivatives of the energy with respect to nuclear positions and nuclear charges ($\frac{\partial^2 E}{\partial Z \partial R} $).
Alchemical forces appear as non-diagonal terms in the generalized Hessian \cite{anm} ( the matrix that contains all the second derivatives of the energy with respect to nuclear charges, atoms' positions, and electron number).
Other off-diagonal terms correspond to nuclear Fukui functions \cite{de1998calculationFukui,baekelandt1996nuclearFukui, balawender2001nuclearFukui,cardenas2020nuclearfukuiforces_ayes} ($\frac{\partial^2 E}{\partial N_e \partial R} $), alchemical Fukui functions \cite{von2006molecular, marcon2007tuning, munoz2017predictive, balawender2019exploringGeerlings, gomez2021alchhardness_cardenas, cardenas2016chem_pot} ($\frac{\partial^2 E}{\partial N_e \partial Z} $)

Those functions have already been studied, while to the best of our knowledge no paper has yet carried out an in-depth analysis of alchemical forces. 
To study the predictive power and applications of alchemical predicted gradients and Hessians, we just consider the restricted Hartree Fock case.
We think that an extension to UHF or DFT is possible, and some of the conclusions derived for RHF may also hold for other levels of theory.

In the first section of this paper, we will introduce the reader to the concept of APDFT, as well as to the Coupled Perturbed Hartree-Fock (CPHF) method used to calculate alchemical derivatives. We will derive an analytical formulation of RHF alchemical forces.
In the second section we show that it is possible to compute higher-order derivatives through numerical differentiation and those derivatives can be used to predict geometrical gradients and Hessians for the target molecules, we also discuss how basis sets affect the accuracy of alchemical predictions of energy and geometry.
In the results section, we give some examples of how to use the predicted gradients and Hessians to relax the geometry structures of the target molecules.

 \section{Methods}
\subsection{\label{sec:level1}Alchemical Perturbation Density Functional Theory}
Within the Born Oppenheimer approximation\cite{Born_Oppenheimer}, the non-degenerate electronic ground state wave function $\Psi_0(\mathbf{r};\mathbf{R},\mathbf{Z},N_e)$ depends parametrically on the positions of the nuclei $\mathbf{R}$, the nuclear charges $\mathbf{Z}$, and the number of electrons $N_e$.
The nuclear charge vector $\mathbf{Z}$ specifies the chemical composition of a given molecule, for real molecules $\mathbf{Z}$ can only have integer values, though from a theoretical point of view it is possible to extend the concept of atomic charge to any real number. 
It is thus possible to define a smooth path for $\mathbf{Z}$ coordinates that connects molecules of different chemical composition \cite{von2006molecular}. We call "alchemical transmutation" the process of transforming one element into another for one or more atoms in a molecule.
In order to connect a reference molecule R to a target molecule T we can define the alchemical coordinate $ \lambda$ as a linear transformation of the nuclear charges from the reference $\mathbf{Z}^\text{R}=\mathbf{Z}(\lambda=0)$ to the target $\mathbf{Z}^\text{T}=\mathbf{Z}(\lambda=1)$.
\begin{equation} \label{lambda_def}
  \begin{aligned}
  \mathbf{Z}(\lambda) &\equiv \mathbf{Z}^\text{R} + \lambda (\mathbf{Z}^\text{T}-\mathbf{Z}^\text{R}) = \mathbf{Z}^\text{R} + \lambda  \Delta \mathbf{Z}
\end{aligned}
\end{equation}
The linear transformation of the nuclear charges corresponds also to a linear transformation of the electronic Hamiltonian and the nuclear-electron attraction operators $\hat{V}_{ne}$:
\begin{equation}
\begin{aligned}
     \hat{H}(\lambda) &= \hat{H}^R + \lambda (\hat{H}^T-\hat{H}^R) = \hat{H}^R + \lambda (\hat{V}_{ne}^T-\hat{V}_{ne}^R) \\
     \hat{H}(\lambda) &= \hat{H}^R + \lambda\Delta \hat{V}_{ne}
\end{aligned}
\end{equation}
For isoelectronic transmutations, the total electronic ground state energy of the molecule is continuous and differentiable in $\lambda$. \cite{Lilienfeld2009} 
From the energy of the reference molecule $E^\text{R} = E(\lambda=0)$ and its derivatives, we can express the energy of the target molecule $E^\text{T}= E(\lambda=1)$ through a Taylor series expansion, where $\Delta\lambda=1$ and where we assume convergence.
\begin{equation} \label {APDFT_taylor}
   E^\text{T} = E^\text{R} + \sum_{n=1}^{\infty}{ \frac{1}{n!} \left.\frac{\partial^n E(\lambda)}{\partial \lambda^n}\right|_{\lambda=0}}
\end{equation}

Truncating the series in Eq.\ref{APDFT_taylor} to a certain order will give an approximation of $E^\text{T}$. We call APDFTn prediction the approximation that includes all terms up to the n$^\textit{th}$ order.

For a variational wave function, the first alchemical derivative of the electronic energy can be evaluated via the Hellmann-Feynman theorem \cite{HF_theorem_Feynman,Lilienfeld2009}:

\begin{equation}
\label{firstderiv}
\begin{aligned}
\left. \frac{\partial E}{\partial \lambda}\right|_{\lambda=0} & = \bra{\psi_\text{R}}\hat{H}_\text{T} - \hat{H}_\text{R} \ket{\psi_\text{R}} \\
& = \int_\Omega d\mathbf{r} \Delta V_{ne}(\mathbf{r}) \rho_\text{R}(\mathbf{r}) \\
\end{aligned}
\end{equation}

For a monodeterminantal wave function (e.g. the solution of an Hartree Fock or a Kohn-Sham DFT calculation) the first alchemical derivative (Equation \ref{firstderiv}) can be rewritten as a contraction between the one electron density matrix $P$ of the reference molecule with the change in the nuclear-electron potential energy operator $\Delta V$:
\begin{equation}
\label{firstderiv_HF}
\begin{aligned}
\left. \frac{\partial E}{\partial \lambda}\right|_{\lambda=0} & = \sum_{\mu\nu}P_{\mu\nu}\Delta V_{\mu\nu} 
\end{aligned}
\end{equation}

Higher order derivatives can be obtained via numerical differentiation or analytically using either M\o{}ller Plesset perturbation theory \cite{Chang_bonds} or a coupled perturbed approach (CPHF or CPKS-DFT) \cite{zachara2012Geerlings}.
The contribution of nuclear nuclear repulsion to the alchemical derivatives can be obtained analytically as the derivative of a classical charge distribution.

\subsection{Coupled Perturbed Hartree Fock}
Coupled perturbed methods are a powerful tool to calculate derivatives properties of a mono determinant wave function (HF,DFT). They represent the standard way to calculate polarization \cite{cammi_pcm, cave1969CPHF, dalgarno1962cphf} and can also be used for alchemical perturbations. \cite{balawender2019exploringGeerlings,c60_2018Geerlings,benzene2013Geerlings, Balawender2018}

The mathematical treatment to calculate the molecular response to an external electric field or to the electric field generated by the transmutation of one atom is indeed identical; here we would like to recall some theoretical aspects of the coupled perturbed Hartree Fock method.

In CPHF the first derivative of the coefficient's matrix $C$ is written as a the product of $C$ and a unitary response matrix $U$:
\begin{equation}
\label{CU_CHPF}
    \left (\frac{\partial C} {\partial \lambda} \right)=CU
\end{equation}

From the orthogonality condition $ C^TSC = I $ it is possible to show that $U$ must be skew symmetric.
\begin{equation}
    \begin{aligned}
        &\frac{\partial}{\partial \lambda}(C^TSC)=0 \\
        &U^T(C^TSC)+(C^TSC)U=0 \\
        &U=-U^T
    \end{aligned}
\end{equation}

Since $U$ acts as a rotation matrix, and since orbital rotations inside the occupied-occupied and the virtual-virtual blocks do not affect the total energy, we can choose the $o-o$ and the $v-v$ blocks of $U$ to be zero; keeping only the $o-v$ blocks of $U$ non zero.
The derivative of the Roothaan equation $FC=SC\epsilon$ with respect to $\lambda$ becomes:
\begin{equation}  \label{CPHF_1}
\frac{\partial F}{\partial \lambda}C+FCU =SCU \epsilon +SC\frac{\partial \epsilon}{\partial\lambda}\\
\end{equation}

Multiplying from the left by $C^T$ and simplifying using the identities $C^T S C= I $ and $C^T F C= \epsilon $, leads to:
\begin{equation}  \label{CPHF_2}
C^T \frac{\partial F}{\partial \lambda}C=U\epsilon -\epsilon U + \frac{\partial \epsilon}{\partial\lambda}
\end{equation} 
Since $\epsilon $  is a diagonal matrix and $U, U \epsilon , \epsilon U$ are zero on the diagonal blocks we can divide Equation \ref{CPHF_2} in two parts:
\begin{align}
    \left(C^T \frac{\partial F}{\partial \lambda}C\right)_{ia} &= U_{ia}(\epsilon_{aa} -\epsilon_{ii}) \tag{out  of  diagonal} \\
    \left(C^T \frac{\partial F}{\partial \lambda}C\right)_{pq}& = \frac{\partial \epsilon_{pq}}{\partial \lambda}
    \tag{on the diagonal}
\end{align}

We used the standard index notation for molecular orbitals, $i,j,k$ for occupied orbitals, $a,b,c$ for virtual orbitals, and $p,q,r$ for arbitrary orbitals. we will use Greek letters $\lambda,\mu,\nu,\sigma$ as indexes of atomic orbitals and capital letters $I,J,k$ as atoms' indexes. 

The partial derivative of the Fock matrix in Equation \ref{CPHF_2} contain the term $\partial V / \partial \lambda$, but also an implicit dependence on $U$ since $F$ depends on $C$. A detailed solution to the CPHF equations can be found in Refs.\cite{pople1979derivative, cammi_pcm, zachara2012Geerlings}
After solving Eq. \ref{CPHF_2} derivatives of the one particle density matrix can be evaluated as:
\begin{equation}
\begin{aligned}
    \frac{\partial P_{\mu\nu}} {\partial \lambda} &= \frac{\partial \left(C_{\mu i}C^T_{i\nu} \right)} {\partial \lambda}\\
    &=C_{\mu a}U_{ai}C_{i\nu}^T+C_{\mu i}U_{ia}^TC_{a\nu}^T 
\end{aligned}
\end{equation}

\subsection{Second and third energy derivatives from atomic contributions}
In APDFT any molecular property can be expressed as a function of the alchemical coordinate $\lambda$. \\
Explicating the dependence on the nuclear charges ($A(\lambda)= A\left(\mathbf{Z} (\lambda)\right)$), differentiation with respect to $\lambda$ can be performed using the chain rule:
\begin{equation} \label{decomposition}
\begin{aligned}
    \frac{\partial^n A} {\partial \lambda^n} &= \left(\sum_I^{atoms} \frac{\partial Z_I}{\partial \lambda} \frac {\partial}{\partial Z_I}  \right)^n A \\
    \frac{\partial Z_I}{\partial \lambda}&= Z_I^T- Z_I^R
\end{aligned}
\end{equation}
The last equation comes directly from Eq.\ref{lambda_def} . 


Solving the CPHF equation for any atom $I$ leads to the evaluation of a response matrix $U^I$ and of the first order derivatives ($ \Delta V^I,  C^I,  F^I,  \epsilon^I$) of the operators with respect to the nuclear charge $Z_I$.

Consistently with the Wigner $2n+1$ rule, from these first order derivatives it is possible to evaluate the second and the third derivatives of the electronic energy:

\begin{equation}\label{2nd_derivCPHF}
\frac{\partial ^2 E}{\partial Z_I \partial Z_J}=     4 \sum_a \sum_i U^I_{ai}\Delta V^J_{ai}
\end{equation}

\begin{equation}
\begin{aligned} \label{3rd_derivCPHF}
\frac{\partial^3 E}{\partial Z_I \partial Z_J \partial Z_K}=4[(IJK)+ (JKI)+(KIJ)]\\
(IJK)=\sum_{i}^{occ.}\sum_{\mu\nu}
F_{\mu \nu}^I C_{\mu i }^J C_{\nu i}^K
-\sum_{ij}^{occ.}\sum_{\mu\nu} S_{\mu\nu} C_{\mu i}^I C_{\nu j}^J \epsilon^K_{ij}
\end{aligned}
\end{equation}

This formulation is particularly convenient for highly symmetrical systems, where many of the derivatives ${\partial A}/{\partial Z_I}$ are symmetrically equivalent.
Only few explicit CPHF calculations are needed, while the number of possible targets grows combinatorially with the size of the system.
\subsection{Basis set energy correction}
\label{sec:bsec}
We pointed out in a previous paper \cite{Domenichini2020} that
neglecting the derivatives of the basis set coefficients with respect to nuclear charges, i.e. calculating alchemical derivatives using the basis set of the reference molecule only, the alchemical series converges to the energy of the target molecule with the basis set of the reference ($E^\text{T[R]}$).
The difference between this energy and the true target energy was named (alchemical) basis set error: $\Delta E_{\textrm{BS}} := E^\text{T[R]} - E^\text{T[T]}$. \\
In the same paper we proposed a correction to the basis set error that can be added to any order APDFT energy prediction. The correction decomposes the total basis set error into a sum of individual atom contributions calculated from isolated atoms.

\begin{equation}
\Delta E_\text{BS}^{correction}:= \sum_{I}^{atoms} E^\text{T[T]}_I-E^\text{T[R]}_I
\end{equation}

It is worth to mention the existence of the universal Gaussian basis set (UGBS)\cite{ugbs}, an uncontracted basis set where all elements share the same basis sets exponents, thus there is no difference between reference and target basis sets.
 \subsection{Alchemical forces} 
Alchemical forces are the mixed derivatives of the energy with respect to both nuclear charges and nuclear positions \cite{anm}. Alchemical forces can be computed analytically either by differentiating the alchemical potential with respect to the nuclear coordinates or differentiating the geometrical gradient with respect to the nuclear charges. In accordance with Schwarz's theorem both methods are valid and will be presented below.
\subsubsection{First formulation}
Differentiating the first alchemical derivative of the electronic energy (Eq.\ref{firstderiv_HF}) with respect to the nuclear coordinates $\mathbf{R}_I$ of the atom $I$  leads to:
 \begin{equation}
     \frac{\partial}{\partial \mathbf{R}_I}\left(P\Delta V^{ne}
     \right)=\frac{\partial P}{\partial \mathbf{R}_I} \Delta V +
    P \frac{\partial \Delta V}{\partial \mathbf{R}_I}     
 \end{equation}
 where:
 \begin{equation}
     \frac{\partial \Delta V}{\partial \mathbf{R}_I}=
     \frac{\partial  V}{\partial \mathbf{R}_I} \frac{\partial Z_I}{\partial \lambda }
 \end{equation}

 The derivative of the one electron density matrix $\frac{\partial P}{\partial R}$ can be obtained through a CPHF calculation, independently for each molecular coordinate $R$, i.e. three times the number of atoms, for this reason we think is more convenient following the other approach.
\subsubsection{Second formulation}
Differentiating the geometrical gradient expression with respect to the nuclear charge of an atom ($Z_I$) requires only one density derivative $\frac{\partial P}{\partial Z_I}$, which is also needed for APDFT3 energy predictions. 
 

The first derivative of the energy with respect to one nuclear Cartesian coordinate $R$ is: \cite{pople1979derivative}
\begin{equation} \label{geom_gradient}
\begin{aligned}
\frac{\partial E}{\partial R}&= \sum_{\mu\nu}P_{\mu\nu}\frac{\partial H_{\mu\nu}^{(1)}}{\partial R}\\
&+\frac{1}{2}\sum_{\mu\nu\lambda\sigma}
P_{\mu\nu}P_{\lambda\sigma}\frac{\partial}{\partial R}(\mu \lambda | | \nu\sigma)\\
&-\sum_{\mu\nu}W_{\mu\nu}\frac{\partial S_{\mu\nu}}{\partial R}
\end{aligned}
\end{equation}
Where $H_{\mu\nu}^{(1)} $ is the mono electronic part of the Hamiltonian,$(\mu \lambda | | \nu\sigma)$ is the sum of the bielectronic Coulomb and Exchange integrals, and $W$ is the energy weighted density matrix:
\begin{equation}\label{W_def} 
\begin{aligned}  
     (\mu \lambda | | \nu\sigma) \coloneqq  & \bra{\phi_\mu(\mathbf{r}_1)\phi_\lambda(\mathbf{r}_2)}
\frac{1}{|\mathbf{r}_1-\mathbf{r}_2|}  \ket{\phi_\nu(\mathbf{r}_1)\phi_\sigma(\mathbf{r}_2)}   - \\  \frac{1}{2}&\bra{\phi_\mu(\mathbf{r}_1)\phi_\lambda(\mathbf{r}_2)}
\frac{1}{|\mathbf{r}_1-\mathbf{r}_2|}  \ket{\phi_\nu(\mathbf{r}_1)\phi_\sigma(\mathbf{r}_2)}   \\
    W_{\mu\nu}\coloneqq & \sum_i ^{occ.} \epsilon_i C_{\mu i} C_{\nu i}^T
\end{aligned}
\end{equation}
Eq.\ref{geom_gradient} is composed of three terms, and we need to differentiate all of them with respect to the nuclear charge $Z_I$ of a given atom $I$. 

For the first term of Eq.\ref{geom_gradient}:
\begin{equation} \label{firts_term_af}
     \frac{\partial}{\partial Z_I} \left( P \frac{\partial H^{(1)}}{\partial R}\right)= 
\frac{\partial P}{\partial Z_I}\frac{\partial H^{(1)}}{\partial R}+ P \frac{\partial^2 H^{(1)}}{\partial Z_I \partial R }
\end{equation}

The mono electronic Hamiltonian $H^{(1)}$ is composed of two parts: the kinetic energy operator $T$, which is independent of $Z_I$, and the nuclear electron potential $V$.
\begin{equation} \label{dv_dZdR}
\begin{aligned}
\frac{\partial^2 H^{(1)}}{\partial Z_I \partial R } &= \frac{\partial^2 (T+V)}{\partial R \partial Z_I}  =\frac{\partial}{\partial R}\frac{\partial V_{\mu\nu}}{\partial Z_I} \\
&=\frac{\partial } {\partial R} \frac{\partial}
{\partial Z_I} \sum_J^{atoms} \bra{\phi_\mu(\mathbf{r})}
\frac{Z_J}{|\mathbf{R}_J-\mathbf{r}|}
\ket{\phi_\nu(\mathbf{r})}
\\
&=\frac{\partial }{\partial R}\bra{\phi_\mu(\mathbf{r})}
\frac{1}{|\mathbf{R}_I-\mathbf{r}|}
\ket{\phi_\nu(\mathbf{r})} 
\end{aligned}
\end{equation} 

The mixed derivative of the one electron Hamiltonian (Eq.\ref{dv_dZdR}) is only non zero if the coordinate $R$ is referred to the position of atom $I$.

For the second term of Eq.\ref{geom_gradient}:
\begin{equation} \label{dEE_dRdz}
\begin{aligned}
\frac{\partial}{\partial Z_I}(P_{\mu\nu}P_{\lambda\sigma}\frac{\partial}{\partial R}(\mu \lambda | | \nu\sigma) )=\\
 P_{\mu\nu}\frac{\partial P_{\lambda\sigma}} {\partial Z_I}\frac{\partial}{\partial R}(\mu \lambda | | \nu\sigma) + \\
 \frac{\partial P_{\mu\nu}}{\partial Z_I}P_{\lambda\sigma}\frac{\partial}{\partial R}(\mu \lambda | | \nu\sigma) 
\end{aligned}
\end{equation}

Differentiating the third term of Eq.\ref{geom_gradient} yields:
\begin{equation}
\frac{\partial}{\partial Z_I} \left( W\frac{\partial S}{\partial R} \right )=
\frac{\partial W}{\partial Z_I}\frac{\partial S}{\partial R}
\end{equation}
The derivative of $W$ can be obtained from Equation \ref{W_def} using the response matrix $U$ and the derivatives of the MO energies $\partial \epsilon / \partial Z_I$ from the solution of the CPHF equation (Eq.\ref{CPHF_2}):

\begin{equation} \frac{\partial W}{\partial Z_I}= \sum_i ^{occ.} \left( \epsilon_i (CU)_{\mu i} C_{\nu i}^T + 
\epsilon_i C_{\mu i} (CU)^T_{\nu i}   +\frac{\partial \epsilon_i}{\partial Z_I} C_{\mu i} C_{\nu i}^T \right) 
\end{equation}

Collecting together all terms, the electronic part of the alchemical force at a RHF level of theory can than be expressed as follows:
\begin{equation} \label{af_formula}
\begin{aligned}
\frac{\partial^2 E}{\partial R \partial Z_I}&= \frac{\partial P}{\partial Z_I}\frac{\partial H^{(1)}}{\partial R}+ P \frac{\partial^2 H^{(1)}}{\partial Z_I \partial R }\\
&+ \frac{\partial P_{\mu\nu}}{\partial Z_I}P_{\lambda\sigma}\frac{\partial}{\partial R}(\mu \lambda | | \nu\sigma) \\
&-\sum_{\mu\nu}\frac{\partial W_{\mu\nu}}{\partial Z_I} \frac{\partial S_{\mu\nu}}{\partial R}    
\end{aligned}
\end{equation}

\subsubsection{Nuclear-nuclear contribution}
The nuclear nuclear contribution to the alchemical force can be evaluated analytically as derivatives of an electrostatic charge distribution:
\begin{equation} \label{nucnuc_af}
\frac{\partial^2 E_{NN}}{\partial Z_J \partial \mathbf{R}_I}=\frac{Z_I (\mathbf{R}_I-\mathbf{R}_J) }{|\mathbf{R}_I-\mathbf{R}_J|^3}(1-\delta_{IJ})+ \sum_Q^{Q\ne I} \frac{Z_Q (\mathbf{R}_I-\mathbf{R}_Q) }{|\mathbf{R}_I-\mathbf{R}_Q|^3}\ \delta_{IJ}
\end{equation}

\section{Numerical Results}
\subsection{Computational details}
All through our work we used a locally modified version of PySCF \cite{pyscf_article}  in which we implemented subroutines for the analytical alchemical derivatives (Eqs. \ref{firstderiv_HF}, \ref{2nd_derivCPHF}, \ref{3rd_derivCPHF}), as well as the alchemical force (Eqs. \ref{af_formula},\ref{nucnuc_af}).
If not specified differently will be used the pcX-2 basis set\cite{pcXbs} for second row elements in conjunction with pc-2 basis set for hydrogen atoms. 
Basis sets' coefficients and exponents were obtained from Basis Set Exchange \cite{Basis_Set_Exchange, Basis_Set_Exchange2, Basis_Set_Exchange3}.
Geometrical optimization in redundant internal coordinates was performed using a locally modified version of PyBerny \cite{Pyberny}, where we  included the transformation of the geometrical Hessian from Cartesian to internal coordinates.
Root mean square deviations of atomic positions were evaluated using the program "RMSD" \cite{rmsd} implementing the Kabsch algorithm \cite{Kabsch:a12999}.
Molecular sketch of benzene's B-N mutants were plotted using RDKit. \cite{rdkit,landrum2013rdkit,rdkit_algorythm}
All Python code used in this work is made available, free of charge on a Zenodo repository. \cite{domenichini_giorgio_2021} 
\subsection{Convergence of gradients and Hessian series}
\label{sec:higher_order_derivatives}
Geometrical gradients and geometrical Hessians can be expanded as an alchemical series resulting into alchemical predictions of gradients and Hessians of the target molecules. 
The first alchemical prediction of the gradient is the alchemical force (Eqs.\ref{af_formula},\ref{nucnuc_af}).

Higher order derivatives of the gradient can be obtained through numerical differentiation of the alchemical force, while differentiation of the geometrical Hessian was only computed numerically.
Derivatives were performed via a central finite difference stencil of 7 points equispaced by $\Delta\lambda=0.1$.

\begin{figure}[ht]
    \centering
    \includegraphics[width=\linewidth]{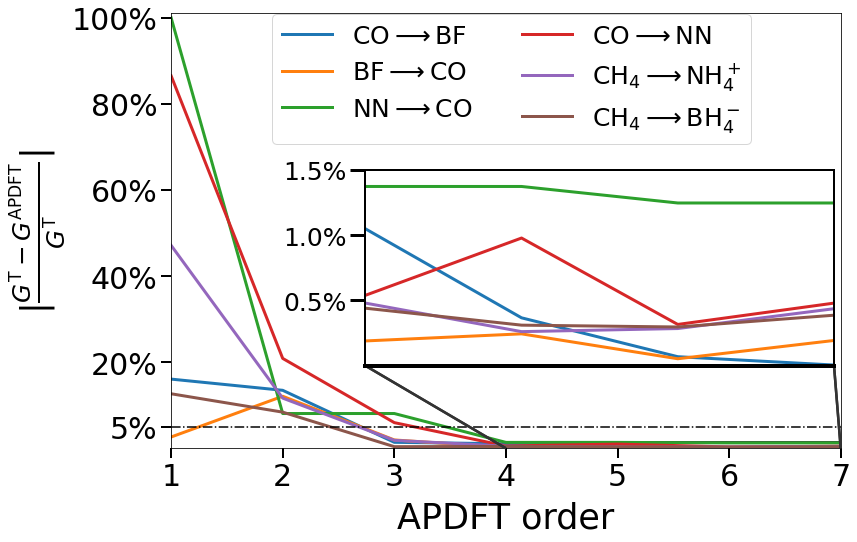}
        \includegraphics[width=\linewidth]{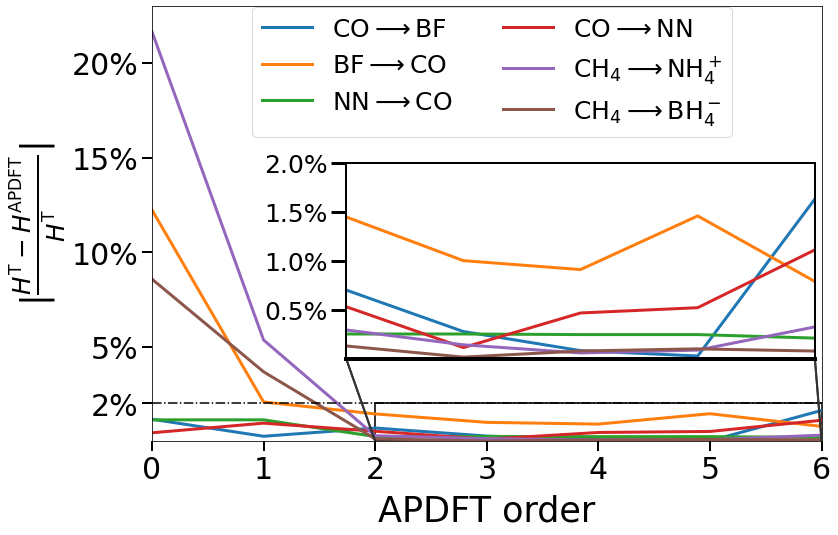}
    \caption{Decreasing error for APDFT predictions of geometrical gradients (TOP) and Hessians (BOTTOM) as perturbation order increases. Legends indicate respective colors for "Ref.~$\rightarrow$ Targ." molecules. Inset shows convergence to less than 1.5\%.}
    \label{fig:error_gradients}
\end{figure}

As a test case we choose to analyse the bond stretching of some diatomic molecules and the stretching of a C-H bond in a methane molecule, the results are plotted in Figure \ref{fig:error_gradients}.

The zeroth order gradient prediction is just the gradient of the reference, which is zero in its geometrical minimum, this leads to a 100\% prediction error for all molecules. Due to symmetry the alchemical force for the transmutation $\text{N}_2\rightarrow$CO is zero, therefore the first order term does not improve the prediction. The inverse transmutation CO$\rightarrow$N$_2$ has also an high first order error which can be justified by similar reasons.

The third order prediction leads to an error of about 5\%, while predictions of orders higher than 4 lead to an error which is in any case below 1.50\%. Going above 5$^\text{th}$ order does not necessarily increase accuracy, because of numerical errors.

Furthermore if the derivatives of the atomic orbitals are neglected the Taylor series converges to the gradient of the target calculated with the basis set of the reference. We can see this effect in the prediction of N$_2\rightarrow$CO where the green line has an offset of approximately 1.3\%.


For Hessians the 0$^\text{th}$ order prediction, i.e. approximating the target Hessian with the reference's one, have still a certain degree of accuracy.
Additional terms increase the precision and going above second order will lead to an error below 2\%.
The best predictions were obtained at the third and fourth and  order, while after the fifth order term the series diverge due to numerical errors in the calculation of the Hessians.
\subsection{Relaxed basis set errors} \label{sec:basis_set_effects}
 
In a previous paper \cite{Domenichini2020} we stated that one of the major sources of error in APDFT vertical energy predictions is the alchemical basis set error, explained in section \ref{sec:bsec}. 

For the sake of this paper is of great interest to analyse the alchemical basis set error error, not only for vertical energy predictions but also for the predictions of energies and geometries of the target in its energy minimum "relaxed errors" ($\Delta E^\text{BS}_{eq},\Delta R^\text{BS}_{eq}$). \\
\begin{equation}
\label{eq:ARBSE}
    \begin{aligned}
    \Delta E^\text{BS}_{eq} := E^\text{T[R]}_{eq} - E^\text{T[T]}_{eq}\\
    \Delta R^\text{BS}_{eq} := R^\text{T[R]}_{eq} - R^\text{T[T]}_{eq}
    \end{aligned}
\end{equation}

In the simple cases of diatomic molecules (BF$\leftrightarrow$CO, CO$\leftrightarrow$N$_2$) at the RHF level of theory, we tested performances of some of the most commonly used the triple and quadruple $\zeta$ basis set.
We compared the family of Karlsruhe def2-nZ \cite{def2tz,def2qz}, the Dunning and coworker's correlation consistent polarized valence cc-pVnZ \cite{Dunning_1989} and polarized core and valence cc-pCVnZ \cite{woon1995a}, the Jensen's polarization consistent pc-n  \cite{pCn_jensen2001a,pCn_jensen2002a,pCn_jensen2007a} and their uncontracted variant optimized for X-ray spectroscopy pcX-n \cite{pcXbs}. 

\begin{figure}[h]
    \centering
    \includegraphics[width=\linewidth]{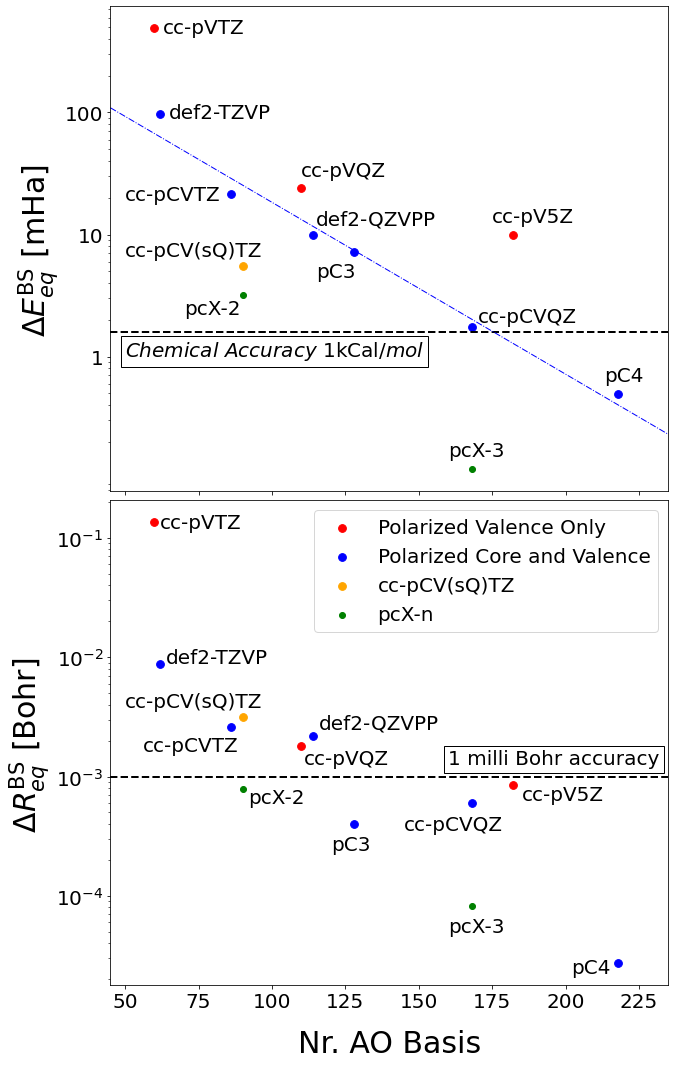}
    \caption{Alchemical basis-set errors (see Eqs.~\ref{eq:ARBSE}) for energy (TOP) and geometry (BOTTOM) as a function of number of atomic basis-functions.
    Values correspond to averages for BF, CO, and N$_2$ (which have been assessed as Ref.~and Targ.~molecules for APDFT predictions in Fig.~2, and Tables I and II). }
    \label{BSerr_comp}
\end{figure} 

Figure \ref{BSerr_comp} shows that $\Delta E^\text{BS}_{eq}$ decrease with the increase in the basis set size, this trend is approximately linear in the logarithmic plot for def2, cc-pCVnZ and pc-n basis sets.
\\ 
Core polarization is important for the alchemical invariance of basis sets; the cc-pVnZ basis functions which are only polarized on the valence shell have a higher $\Delta E^\text{BS}_{eq}$ compared to other basis of equal size. \\
As a proof of this concept we created a new basis set combining the core basis functions (the orbitals with $s$ symmetry) of the cc-pCVQZ basis set with the valence and polarization orbitals of the smaller cc-pCVTZ; the orthonormality of this basis set labeled cc-pCV(sQ)TZ is guaranteed by spherical symmetry. 

In Figure \ref{BSerr_comp} we can compare $\Delta E^\text{BS}_{eq}$ for cc-pCVTZ and cc-pCV(sQ)TZ basis sets: passing from 6 to 8 s type orbitals per atoms is enough to reduce the error by about one order of magnitude.

The cc-pVTZ basis set has a large geometry error, that's because the systems, in order to compensate for the high energy error, increase the orbital overlap between the atoms shortening the bond lengths.
The other basis sets of the cc-pVnZ family (cc-pVQZ and cc-pV5Z) don't suffer as much from this problem.
$\Delta R^\text{BS}_{eq}$ is similar for cc-pCV(sQ)TZ and cc-pCVTZ, both basis set share the same valence and polarization functions.
We can conclude that a good description of core and valence atomic orbitals is needed for accurate APDFT energies, while a good description in valence orbitals and the presence of polarization functions is required for accurate geometry predictions.

Best results in terms of accuracy per basis sets number were achieved both for energy and geometry by the polarization consistent pcX-2 and pcX-3 basis sets. Those are uncontracted basis sets whose coefficients are optimized for X-Ray spectroscopy.
To describe effectively a core-ionization of an atom, the basis functions used should provide a balanced representation of both the ground and core-excited state, where the effective nuclear charge changes by roughly $+1$. The same requirement is also true for alchemy, where the true nuclear charges change by $\pm 1$. 
This is a case where the computational solution to one specific problem (x-rays spectroscopy) can also be used effectively in a completely different application (APDFT).
Acknowledged these conclusions on basis sets, we will use for the rest of the article the pCX-2 basis set, and since the pcX-2 basis sets are defined only for second and third row elements, for hydrogens we will use the pc-2 basis set. 

\section{Applications to geometry relaxation}
In section \ref{sec:higher_order_derivatives} we showed how alchemical perturbation can be used to predict geometrical gradients and Hessians. 
Those gradients and Hessians can be used to relax the geometry of the target molecule in several ways.
We would like to show in this section that a geometrical relaxation using the alchemical predicted gradients and Hessians it is possible and can be accurate depending on the APDFT order.

\subsection{Relaxation techniques}
\label{sec:Optimization_tecniques}
 
The simplest relaxation technique is a single step in the Newton-Raphson optimization method.
The potential energy surface is approximated with a paraboloid; given a a set of Cartesian coordinates $\mathbf{x}$, a gradient $\mathbf{g}$ and a Hessian $H$, the optimization step and the relaxation energy are: 
\begin{align}
    \label{eq:NR_geom_coords}
    \Delta \mathbf{x} = -H^{(-1)} \mathbf{g} \\ \label{eq:NR_energy}
    \Delta E = - \frac{1}{2} \Delta \mathbf{x} H \Delta \mathbf{x} 
\end{align}

A better representation for the monodimensional potential of a bond stretching was given by Morse \cite{Morse_1929}:
\begin{equation}
\label{eq:Morse_Potential}
    V_{Morse}(R;D_e,R_e,a,V_e)= D_e (1-e^{-a(R-R_e)})^2+V_e
\end{equation}

Morse potentials depend on four parameters: the bond dissociation energy  $D_e$, the equilibrium distance $R_e$, the parameter $a$ which controls the width of the potential well, and $V_e=V(r_e)$ the value of the energy at the minimum.

Knowing the energy, the first and second derivatives at a given point of the curve, we can  match the number of parameters with the number conditions using an empirical value for the dissociation energy.
This is legit because the the interpolated curves do not depend sensibly on $D_e$, and analogously to what is done in the UFF \cite{UFF} we will approximate $D_e$ with the value of 100kCal/mol times the total bond order (for UFF this parameter is 80kCal/mol).

For molecules bigger than diatomics is still possible to introduce the Morse potential interpolation as a correction to a Newton-Raphson step.

Using internal coordinates $\mathbf{q}$, we calculated the correction treating every bond if $b$ if it were an independent coordinate:

\begin{equation}
\label{eq:Morse_as_correction1}
    \Delta \mathbf{q}_b^{Corr.} = \Delta \mathbf{q}_b^{Morse}(\mathbf{q}_b,\mathbf{g}_b,H_{bb},D_{e(b)}) -(-\mathbf{g}_b/H_{bb})
\end{equation} 

The mixed terms between different coordinates are still calculated in the paraboloid (Newton Raphson) approximation.

The Morse potential correction affect also the energy:
\begin{equation}
\label{eq:Morse_as_correction2}
\begin{aligned}
    \Delta E^{Corr.}=&-\frac{1}{2}\Delta\mathbf{q}^TH\Delta\mathbf{q} \  \\
    &+\sum_b (\Delta E^{Morse}_b+\frac{1}{2} H_{bb}(\Delta\mathbf{q}_b^{Morse})^2)
\end{aligned}
\end{equation}

A more in detail discussion of the Morse potential interpolation is given in the supplementary materials of this paper.

\begin{figure}
    \centering
    \includegraphics[width=\linewidth]{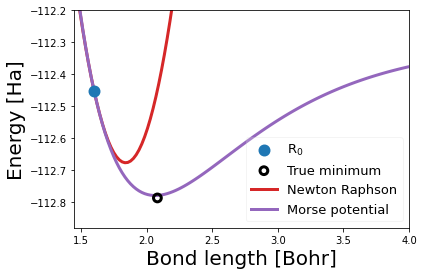}
    \caption{Morse interpolation improves  geometry relaxation as exemplified for CO: 
    Interpolating Hartree-Fock energies, gradients and Hessians at R$_0$ to Morse potential and parabola (Newton Raphson step) affords purple and red potentials, respectively. 
    }
    \label{fig:morse_vs_NR}
\end{figure}

Taking as an example the dissociation of CO, shown in figure \ref{fig:morse_vs_NR}, we can see that the quadratic approximation leads to a poor result when the starting point is far from the minimum. Since the geometries of reference and target are usually substantially different, this condition occurs quite often in alchemical relaxation. 

At the Morse potential describes the whole energy curve of bond dissociations better, and this can lead to a more accurate relaxation, if the gradient and the Hessian are predicted accurately.

\subsection {Diatomics } \label{sec:diatomics}
In the simple case of diatomic molecules we can derive some considerations regarding the accuracy of the alchemical relaxation at different perturbation order, and compare the relaxations obtained using Newton Raphson step and the Morse potential interpolation.

 Tab.\ref{tab:NR_diatomics} shows the results obtained with a NR step. In the predictions  CO$\rightarrow$BF and N$_2\rightarrow$BF the error in bond lengths is consistent, and a higher alchemical order does not give better results.
 
 Furthermore in this method is neglected the dependence of Hessian on the geometry, the different geometries of reference and target lead to a huge error in the prediction of harmonic vibrational frequencies. 
\begin{table}
\centering
\begin{tabular}{*{2}{>{\raggedright\arraybackslash}p{0.12\linewidth}}>{\centering\arraybackslash}p{0.15\linewidth}|*{3}{>{\centering\arraybackslash}p{0.18\linewidth}}}
 Ref.&Targ.& APDFT &   R$_{eq}$ [Bohr] &  $E$[Ha.]  &  $\omega_0$ [cm$^{-1}$] \\\hline \hline 
\multicolumn{6}{c}{Newton Raphson optimization } \\ 
\hline N$_2$   &BF & 2 & 2.290 & -124.2586 & 2676 \\
N$_2$   &BF & 4 & 2.221 & -124.1884 & 2678 \\
CO      &BF & 2  &   2.285 &   -124.1862 &    2402 \\
CO      &BF & 3  &   2.262 &   -124.1598 &    2407 \\
CO      &BF & 4  &   2.258 &   -124.1634 &    2410 \\
True    &BF & -  &  2.353 &   -124.1624 &    1507\\ \hline  
BF     & CO &2 &   1.793 &   -112.7944 &    1415 \\
BF     & CO &3 &   1.846 &   -112.8141 &    1418 \\
BF     & CO &4 &   1.864 &   -112.8121 &    1431 \\
N$_2$  & CO &2 &      2.080 &   -112.7909 & 2740 \\
N$_2$  & CO &4 &      2.076 &   -112.7855 & 2740 \\
True    & CO &-& 2.083 &      -112.7866 & 2430 \\
\hline
BF & N$_2$ & 2  &  0.813 &-109.4238 & 1244 \\
BF & N$_2$ & 3  &  1.375 &-109.1547 & 1274 \\
BF & N$_2$ & 4  &  1.761 &-109.0494 & 1490 \\
CO & N$_2$ & 2  &  1.989 &      -108.9803 &    2416 \\
CO & N$_2$ & 3  &  2.009 &      -108.9977 &    2411 \\
CO & N$_2$ & 4  &  2.005 &      -108.9933 &    2415 \\
True &N$_2$ &- &  2.014 &   -108.9891 &    2730 \\
\hline
MAE &$\Delta Z\pm 1$ & 2 & 0.097 &0.0111&  634 \\ 
MAE &$\Delta Z\pm 1$ & 3 & 0.084 &0.0108 & 635 \\ 
MAE &$\Delta Z\pm 1$ & 4 &0.083 &0.0080 & 632  \\ 
\hline \hline \multicolumn{6}{c}{Morse optimization } \\ \hline
N$_2$&BF& 2  & 2.543 & -124.2904 & 1233 \\
N$_2$&BF& 4  & 2.369 & -124.2033 & 1440 \\
CO &  BF & 2 &  2.412 &  -124.1965 &    1379 \\
CO &  BF & 3 &  2.364 &  -124.1674 &    1453 \\
CO &  BF & 4 &  2.354 &  -124.1704 &    1471 \\
True& BF & - &  2.353 &  -124.1624 &    1507 \\
\hline
BF  &CO& 2 &  2.096 &  -112.7580 &    2744 \\
BF  &CO& 3 &  2.101 &  -112.7867 &    2589 \\
BF  &CO& 4 &  2.104 &  -112.7868 &    2572 \\
N$_2$ &CO& 2 &  2.090 & -112.7913 & 2389 \\
N$_2$ &CO& 4 &  2.084 & -112.7858 & 2408 \\
True &CO& - &    2.083 &     -112.7866 & 2430 \\
\hline
BF & N$_2$  & 2 &   2.133 &       -108.7587 & 5082 \\
BF & N$_2$  & 3 &   2.084 &       -109.0103 & 3510 \\
BF & N$_2$  & 4 &   2.110 &       -109.0010 & 3130 \\
CO& N$_2$ &2 &  2.005 &     -108.9794 &    2910 \\
CO& N$_2$ &3 &  2.019 &     -108.9973 &    2785 \\
CO& N$_2$ &4 &  2.017 &     -108.9929 &    2812 \\
True & N$_2$ & -&  2.014 &     -108.9891 & 2730 \\
\hline
MAE &$\Delta Z\pm 1$ & 2 & 0.022 &  0.0192 & 166 \\ 
MAE &$\Delta Z\pm 1$ & 3 & 0.011 & 0.0045 &  77 \\ 
MAE &$\Delta Z\pm 1$ & 4 &0.007 & 0.0032& 70 \\ \hline
\end{tabular}
\caption{APDFT based predictions of equilibrium distances and energies and frequencies using a single step of the parabola Newton Raphson (TOP) and Morse potential (BOTTOM) optimization.
'True' corresponds to Hartree-Fock SCF results. 
}
\label{tab:NR_diatomics}
\end{table}

Tab.\ref{tab:NR_diatomics} shows that the Morse method performs better than Newton Raphson's.
The biggest errors were made in the predictions N$_2\leftrightarrow$CO where is involved an alchemical change of $\Delta Z =\pm 2$.
All the other predictions which have $\Delta Z =\pm 1$ have an error in the bond length prediction on average of 0.011 Bohr at third order which decreases to 0.007 Bohr at fourth order.
The error in the energy is also very low, 4.5 mHa and 3.2 mHa for APDFT3 and APDFT4 predictions respectively.
The error in the harmonic frequencies prediction is about 70 cm$^{-1}$; is not accurate, but  it is indicative of the value of the frequency. 
\subsection{Alchemical protonation and deprotonation}  \label{sec:deprotonation}
\begin{figure}[h]
    \centering
    \includegraphics[width=\linewidth]
    {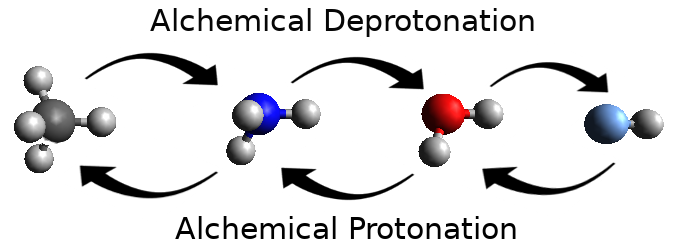}
    \caption{Subsequent alchemical couplings connect the entire iso-electronic series of second period elements' hydrides (CH$_4$,NH$_3$,BH$_2$,HF) through successive alchemical deprotonation (to the right) and protonation (to the left). }
    \label{fig:CH4_serie}
\end{figure}
Deprotonation and protonation energies are important quantities, because describe the changes in enthalpy within an acid-base reactions, and are needed in the calculation of the equilibrium constants p$K_a$ and p$K_b$. \cite{atkins2006response, carvalho2021accurate, pka_CASASNOVAS20095}
In this section, we demonstrate how APDFT can be used to predict protonation and deprotonation energies accurately.
APDFT3 error for vertical deprotonation energies can be as small as 1.4 kcal/mol \cite{Rudorff2020,Cardenas2020Deprotonation, Chang_bonds,Moreno_ADFT}, we would like to go beyond the fixed geometry approximation including the geometrical relaxation energy. More specifically, we exemplify our approach for the alchemical navigation of the iso-electronic 10 electron series of second row hydrides CH$_4 \rightarrow$NH$_3 \rightarrow$H$_2$O$\rightarrow$ HF,(see figure \ref{fig:CH4_serie} for illustration).
Moving from one molecule to a neighbouring one, within alchemical deprotonation (protonation) a hydrogen nucleus is alchemically annihilated (created) while the nuclear charge of the heavy atom is simultaneously increased (decreased).

The initial guess of the protonation site was chose on the electrostatic potential minimum of the reference molecule; in that position we placed a basis set for the hydrogen created.

We implemented a three point finite difference stencil ($\Delta \lambda=0.1$), combining numerical and analytical differentiation we obtained second order Hessian predictions, third order gradient prediction and fifth order energy prediction.

Combining numerical differentiation from a three points finite difference scheme ($\Delta \lambda=0.1$) with analytical differentiation (Eqs.\ref{2nd_derivCPHF}, \ref{3rd_derivCPHF}, \ref{af_formula}), we calculated second order predictions for the Hessian, third order prediction for the gradient and fifth order prediction for the energy.

Geometry relaxation was performed using the Morse potential interpolation (Eqs. \ref{eq:Morse_as_correction1}, \ref{eq:Morse_as_correction2}).

\begin{figure}
    \centering
    \includegraphics[width=\linewidth]{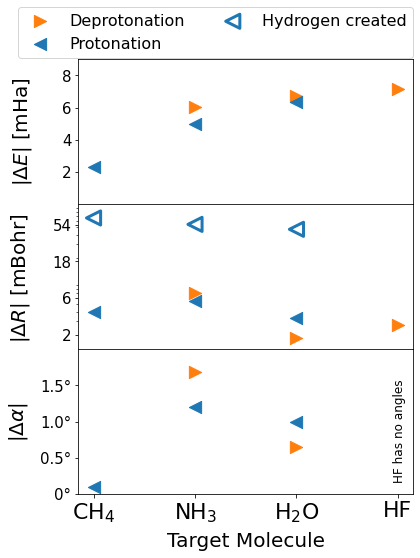}
    \caption{Absolute APDFT prediction errors of equilibrium energies (TOP), bond lengths (MID), and angle widths (BOTTOM)  for the alchemical protonations and deprotonations (see Fig.\ref{fig:CH4_serie}).}
    \label{fig:protonation_deprotonation_results}
\end{figure}

Figure \ref{fig:protonation_deprotonation_results} shows the errors in protonations and deprotonations.
The average error on energy predictions is 7 mHa for deprotonation and 5 mHa for protonation; this error is already present in the vertical predictions, as such does not come from the geometrical relaxations.

Alchemical deprotonation can predict relaxed bond lengths accurately; results show a mean absolute error of 4 milli Bohr and in all cases, the error is lower than 10 milli Bohr, for protonations, the predicted bond length of the hydrogens that were already present in the reference molecule is different from the one of the alchemically created hydrogen.

The effect of alchemical perturbation is much higher on the created hydrogens, for this reason, the error in bond length predictions is  10 times higher than the error for hydrogens already present in the molecule, 56 and 4 milli Bohr respectively.

Angles are more flexible coordinates than bonds, therefore the prediction error of about 1$^{\circ}$ is quite substantial, an exception is the prediction NH$_3\rightarrow$ CH$_4$ where the smaller error might be due to a stiffer energy potential.  
\subsection{B-N Doping of Benzene}

\begin{figure}[h]
    \centering
    \includegraphics[width=\linewidth]{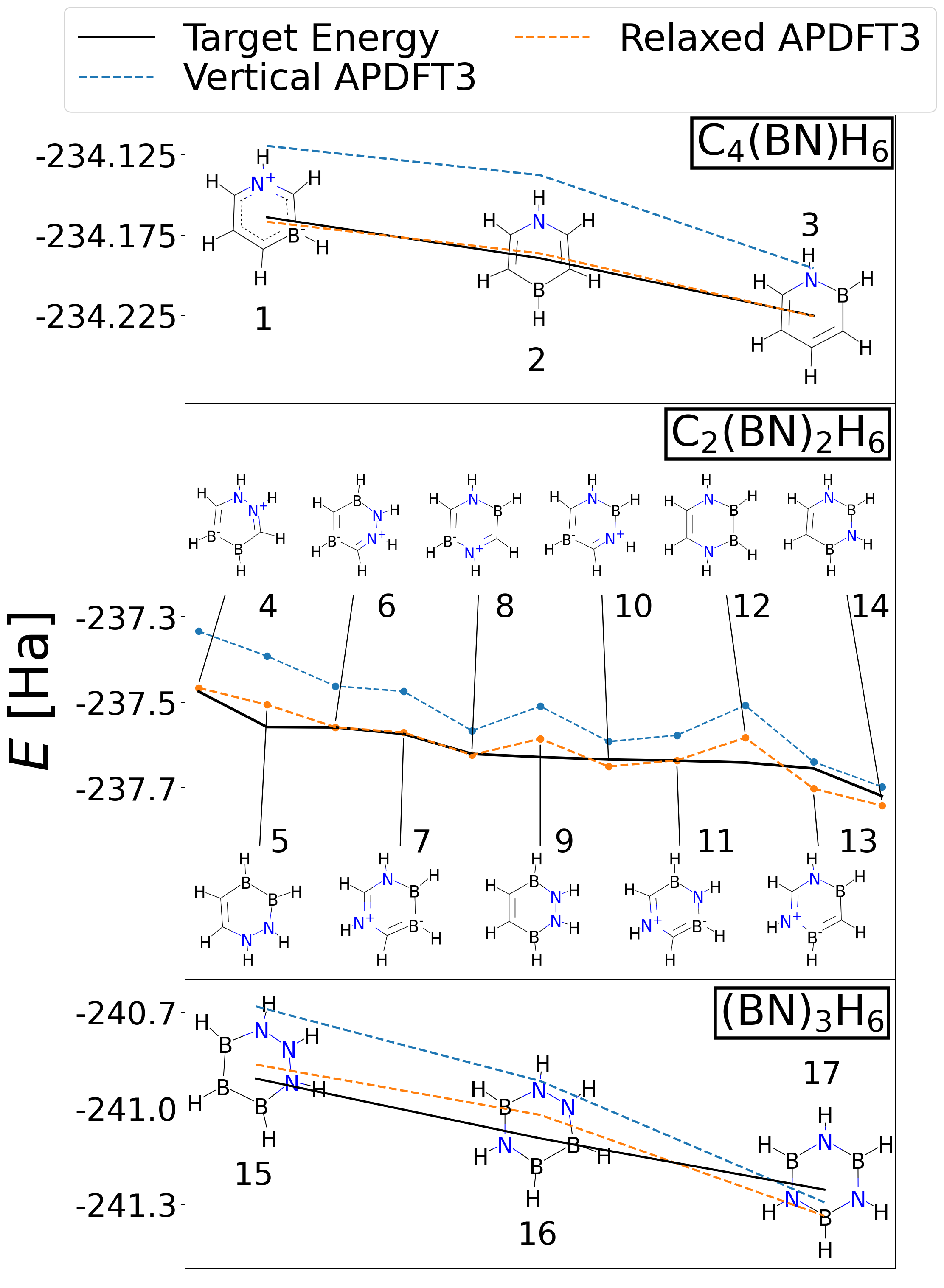}
    \caption{A single APDFT3 reference calculation for benzene affords equilibrium energy predictions for all seventeen possible iso-electronic charge-neutral B-N doped mutants at Hartree-Fock level of theory. }
    \label{fig:benzene_energies}
\end{figure}

APDFT is efficient in predicting properties for multiple targets from a few derivatives when symmetric systems are chosen as reference. An appealing application is the B-N doping of aromatic hydrocarbons such as coronenes, fullerenes, or graphene. \cite{benzene2013Geerlings,anm,Lilienfelds_bn_doped_graphene,c60_2018Geerlings}

Here, we show that is possible to obtain approximate equilibrium energies and relaxed structures for all the 17 B-N doped mutants shown in  Fig.\ref{fig:benzene_energies} through the explicit calculation of just one CPHF derivative for benzene. The CPHF response matrix $U$ can be obtained for all atoms \textit{via} symmetry operations, in particular, we rotated the $U$ matrix around the symmetry axis of the molecule using the Wigner D-matrix \cite{wigner1931gruppentheorie}.
Energy prediction up to third-order can be obtained using Eqs.\ref{firstderiv_HF}, \ref{2nd_derivCPHF}, \ref{3rd_derivCPHF}, the same CPHF derivatives can be reused to calculate the alchemical force in  Eq.\ref{af_formula}. This is sufficient to give a first-order prediction of the geometrical gradient for all 17 target molecules.
The targets' Hessians were approximated at the zeroth order with the Hessian of the reference. 
As shown in Fig.\ref{fig:error_gradients} this approximation has an error of approximately 20\% for the isolectronic hydride series, 
which can still lead to qualitative improvements. In this case, the geometrical relaxation was performed using a Newton Raphson step and not Morse, because it is less sensible to errors in gradient and Hessian.
We measure geometry prediction errors through the root mean square deviation from the true target molecules' minima (RMSD\cite{Kabsch:a12999}).
It is important to remark on the energy ordering of the mutants.
Every B-N substitution corresponds to a decrease in energy of about 3 Hartree, therefore stoichiometry is the dominating dimension.
Among constitutional isomers, energy ordering depends on the relative positions of B and N; for mono doped mutants, the least stable is the one with B and N in \textit{meta} position, followed by the \textit{trans} and the \textit{ortho} isomers.
This ordering is analogous to the reactivity in electrophilic-nucleophilic aromatic substitutions \cite{aromatic_substitutions}. 
In fact, after a B-N doping of benzene, the electronic charge shifts from the boron atom to the nitrogen; in this context nitrogen atoms act as Lewis acids (electrophiles) and borons as Lewis bases (nucleophiles). \cite{robles2018localelectroph,Ayers_electrophilic_2005} 
We note, however, that these trends change upon removal of the nuclear repulsion terms which do not affect the electronics.
The predicted energy ordering is the same, both for the vertical and the relaxed energies, the relaxation calculated only with the first order alchemical force is in fact not able to discriminate alchemical enantiomers \cite{guido2021AlchChirality} and does not change the energy order.

\begin{figure}[h!]
    \centering
    \includegraphics[width=\linewidth]
    {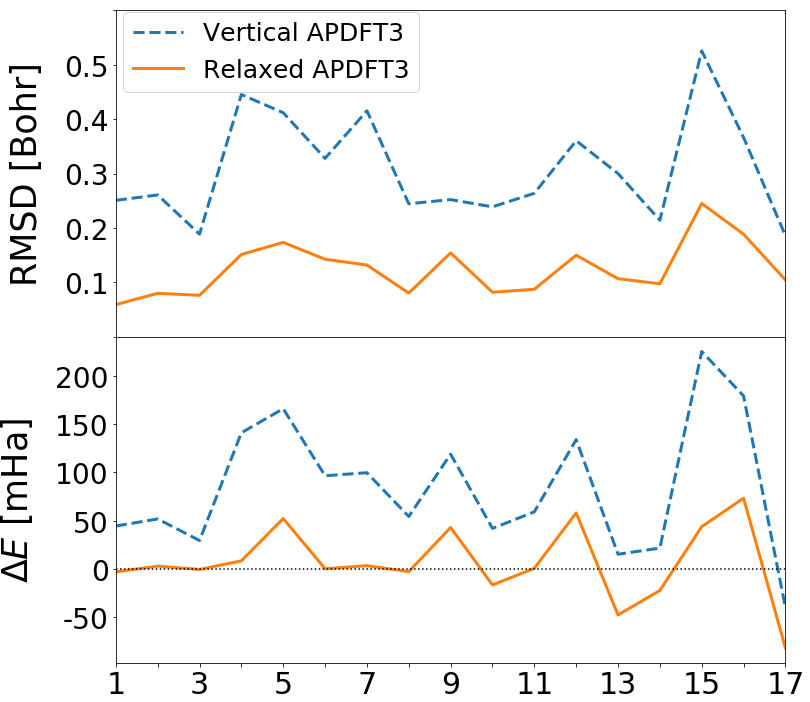}
    \caption{Geometry (TOP) and equilibrium energy (BOTTOM) prediction errors from a single APDFT3 reference calculation for benzene for all seventeen possible iso-electronic charge-neutral B-N doped mutants (structures shown in Fig.\ref{fig:benzene_energies}).}
    \label{fig:benzene_pred_error}
\end{figure}


Relaxation reduces the average RMSD by a factor of 60\% : from 0.31 Bohr to 0.12 Bohr. For singly doped mutants the RMSD is reduced from 0.23 Bohr to only 0.07 Bohr.

The highest deviation from benzene's geometry was found for mutants with adjacent N-N atoms and adjacent B-B atoms (mutants 4,5 7,15), the lowest for the mutants with alternating B-N (mutants 3,14,17). 
This is a consequence of alchemical symmetry \cite{guido2021AlchChirality}: at first order, the alchemical force on a B-N bond is zero, and consecutively there will be no stretching.

The mean absolute error in energy predictions is reduced by 70\% after geometrical relaxation, passing from 89 mHa to 27 mHa. In particular, for singly B-N doped mutants it is reduced from 42 mHa to just 2 mHa.

\section{Conclusion}
In this work, we have shown that alchemical perturbation can successfully be applied to geometry relaxation and the prediction of equilibrium energies.
A key step is the computation of the alchemical force ($\frac{\partial^2 E}{\partial Z \partial R} $), for which we gave an analytical formulation in the Restricted Hartree Fock case, an extension of the formula to other self consistent field methods such as UHF and DFT should be possible.
As we reported previously for vertical APDFT predictions \cite{Domenichini2020}, the basis set choice is crucial for accuracy. We confirm this conclusion also for the APDFT based geometry relaxation and equilibrium energies studied in this work. More specifically, using the pcX-2 basis set, gradients, and Hessian can be predicted with a 4$^\text{th}$ order error smaller than 2\%.
The consecutive geometrical relaxation using the employed Morse potential interpolation can predict equilibrium energies and bond lengths with a respective accuracy of $\sim$10 mHa and $\sim$0.01 Bohr for small molecules (BF, CO, N$_2$, and CH$_4$, NH$_3$, H$_2$O, and HF), and $\sim$27 mHa and $\sim$0.12 Bohr for all the BN doped benzene mutants. 

In comparison to vertical APDFT predictions, 
we have confirmed the expectation that the inclusion of gradient and Hessian information results
in a considerable increase in accuracy. 
The computational overhead is modest 
since first order APDFT predictions of gradients do not require more CPHF calculations than the one required for APDFT3. 
As for all alchemical methods, APDFT based geometry relaxations become particularly attractive when symmetry reduces number derivatives and simultaneously increases number of target molecules. 
In particular, we demonstrated reasonable equilibrium energy and geometry predictions for all the seventeen BN doped mutant systems based on a single APDFT3 reference calculation for benzene. 
As an outlook, we believe that our findings and discussion would suggest that APDFT based geometry relaxation can be meaningful for many compounds on regular lattices, e.g.~doped fullerenes or graphitic systems.

\section{Acknowledgement}
We acknowledge support from the European Research Council (ERC-CoG grant QML and H2020 project BIG-MAP). This project has received funding from the European Union's Horizon 2020 research and
innovation program under Grant Agreement \#772834
and \#957189. All results only reflect the
authors' view and the EU is not responsible for any use that may be made of the information. This work was partly supported by the NCCR MARVEL, funded by the Swiss National Science Foundation. The computational results presented have been achieved
using the Vienna Scientific Cluster (VSC).
The authors would also like to thank Dr.~Max Schwilk for the helpful discussions about the formulation of analytic derivatives and Prof.~Frank Jensen from Aarhus University for the suggestion of using the pc-X  basis set in the context of quantum alchemy.

\section{References}
\bibliography{bib_files/BasisRef,bib_files/Alchemy_vLg,bib_files/Alchemy_Others, bib_files/Alchemy_Cardenas, bib_files/Alchemy_Geerlings, bib_files/Alchemy_Keith, bib_files/software, bib_files/Other_works ,bib_files/QML, bib_files/derivatives,bib_files/Geomopt.bib}

\end{document}


\title{Alchemical geometry relaxation - Supplementary materials}
\author{Giorgio Domenichini and O.Anatole von Lilienfeld}
\email{anatole.vonlilienfeld@univie.ac.at}
\maketitle

\section{Supplementary Tables}
\begin{table}[h!]
\setlength{\tabcolsep}{10pt}
\renewcommand{\arraystretch}{1.5}
    \begin{tabular*}{\linewidth}{l@{\extracolsep{\fill}}|ccc}
    \toprule
    & CO$\rightarrow$N$_2$ & BF$\rightarrow$CO & CH$_4\rightarrow$NH$_4^+$\\
    \hline 
    \\[-3ex]
\midrule
    CFD $O(\Delta \lambda^2)$  \ \ \ \  & 0.224238  &   0.266513    &   0.078847  \\
    CFD $O(\Delta \lambda^4)$     & 0.224561  &  0.266778    &   0.078920 \\
    CFD $O(\Delta \lambda^6)$    & 0.224562  &  0.266777    &  0.078920 \\
    Analytical                  & 0.224556  &   0.266777    &  0.078919    
    \end{tabular*}
    \caption{Comparison between the values of $\partial^2 E / \partial x \partial\lambda$ (atomic units: Hartree*Bohr$^{-1}*e^{-1}$) calculated using the analytical formula proposed and the numerical differentiation of the geometrical gradient. For the central finite difference stencil was chosen a spacing $\Delta \lambda$=0.1. The geometrical coordinate $x$ indicates the bond stretching C-O,B-F and C-H.}
    \label{tab:comparison_against_fd}
\end{table}

\begin{table}[h]
\centering
\begin{tabular}{*{2}{>{\arraybackslash}p{0.1\linewidth}}|>{\centering\arraybackslash}p{0.3\linewidth}>{\centering\arraybackslash}p{0.2\linewidth} >{\raggedleft\arraybackslash}p{0.18\linewidth}}
Ref. &Targ. & Bond A-H [Bohr] & Angle $\hat{\mathrm{HAH}}$ &  Energy\\ \hline
true & CH$_4$ & 2.0439 &109.5°&-40.2152 \\  \hline
CH$_4  $ & NH$_3$ & 1.8785 &109.9°& -56.2282\\
true & NH$_3$ & 1.8855 & 108.2° &-56.2222 \\  \hline
NH$_3  $ & H$_2$O & 1.7774 &107.0°&-76.0701\\
true & H$_2$O& 1.7756 & 106.4° & -76.0633\\ \hline
H$_2$O &  HF& 1.6931 & - & -100.0727\\
true & HF & 1.6957 & - & -100.0655 \\ \hline
MAE& \  & 0.0038 & 1.2° & 0.0066
\end{tabular}
\caption{Alchemical deprotonation predictions and true values of energy and geometrical parameters of second row hydrides. Bonds and angles among hydrogens and the central atom are all equivalent due to symmetry, before and after alchemical deprotonation.}
\label{tab:deprotonation}
\end{table}

\begin{table}[h]
    \centering
    \begin{tabular}{*{2}{>{\arraybackslash}p{0.1\linewidth}}|>{\centering\arraybackslash}p{0.3\linewidth}>{\centering\arraybackslash}p{0.2\linewidth} >{\raggedleft\arraybackslash}p{0.18\linewidth}}
    Ref. &Targ. & Bond A-H [Bohr] & Angle $\hat{\mathrm{HAH}}$ &  Energy\\ \hline
true & HF & 1.696 & - & -100.066 \\ \hline
HF & H$_2$O& 1.772,1.728 & 107.4° & -76.057\\
true & H$_2$O& 1.776 & 106.4° & -76.063\\ \hline 
H$_2$O & NH$_3 $& 1.880,1.831 & 109.4°,109.3°& -56.217\\
true & NH$_3$ & 1.886 & 108.2° &-56.222 \\ \hline
 NH$_3$ & CH$_4$ & 2.048,1.978 & 109.4°,109.6°& -40.213\\
true & CH$_4$ & 2.044 &109.5°&-40.216 \\ \hline
MAE &  &0.004,0.056 & 0.8°,0.6° & 0.0046 \\

\end{tabular}
    \caption{Energy and geometry predictions from alchemical protonation. The two values for bond lengths and angle widths are referred to the hydrogens already present and to the grown up hydrogen respectively. }
    \label{tab:protonation}
\end{table}

\section{Supplementary method: Morse potential interpolation}

\subsection{The Morse potential}
The Morse potential is a good approximation of the binding potential of a diatomic molecule. 
\begin{equation}
\label{Eq:Morse_Potential}
    V_{Morse}(r;D_e,a,r_e,V_e)= D_e (1-e^{-a(r-r_e)})^2 +V_e
\end{equation} 

The Morse potential is a function of $r$, the bond length, and its shape depends on four parameters: $D_e$ the positive defined Dissociation Energy, $r_e$ the distance of equilibrium, the parameter $a$ that gives the width of the energy hole and $V_e= V(r_e)$ is the energy at equilibrium.\\

Knowing three conditions: value of energy, gradient and Hessian in a given point, it is not sufficient to uniquely identify a Morse potential. Though we will show that approximated dissociation energy can lead to Morse potentials which are accurate enough. 

Defining $t(r):=e^{-a(r-r_e)}$ and using the derivation rule: $dt/dr=-at $ we can build the system of equations:
 \begin{equation} 
 \label{Eq:Morse_system}
\begin{cases} 
V= D_e (1-t)^2 +V_e \\ \\
g= 2D_e a t(1-t)  \\ \\
H = 2D_e a^2t(2t-1)
\end{cases}
\end{equation}
 
Combining the second and the third term of this system is possible to simplify $a$ and find an expression which depends uniquely on $t$.

\begin{equation}
\frac{g ^2}{2H D_e }=Z= \frac{ t(1-t)^2}{2t-1}    
\end{equation} 

This expression can be arranged in a third order equation in $t$.
\begin{equation}
t^3-2t^2+(1-2Z)t-Z=0
\end{equation}

There are three analytical solutions for $t$. 
Since $t$ is defined as an exponential needs to be always greater than zero. 
Since $a,D_e$ are positive defined quantities, from the second equation of the system we can deduce that $t(1-t)$ needs to be of the same sign of $g$.
\begin{equation}
\begin{cases}
0<t<1 \text{\ \   if \ \  } g>0  \\ \\
 1<t \text{\ \   if \ \ } g<0 
\end{cases}
\end{equation}
Once $t$ is found $a$, $r_e$,$V_e$ can be obtained through the relationship:
\begin{equation}
\begin{aligned} 
& a=\frac{g}{2D_e t(1-t)} \\ \\
&r_e=\frac{\log{t}}{a}-r\\ \\
&V_e=V-D_e(1-t)^2
\end{aligned}
\end{equation}

\subsection{Harmonic vibrational frequencies}
From a Morse potential we can approximate the vibrational frequency or \textit{via} the Harmonic approximation or \textit{via} the quantum states of the Morse potential (anharmonic). 

Is possible to get the second derivative in the minimum replacing $t=1$ in the last term of Eq.\ref{Eq:Morse_system}  
\begin{equation}
    k_e = 2D_e a^2
\end{equation}
To obtain the wave number $\tilde \nu$:
\begin{equation}
\begin{aligned}
h \nu=\sqrt{\frac{k_e}{\mu}} \\
\tilde{\nu} = \frac{\nu}{c}  = \frac{1}{hc} \sqrt{\frac{k_e}{\mu}}
\end{aligned}
\end{equation}
\subsection{In the limit of $D_e\rightarrow\infty $}
We want to solve the system in Eq. \ref{Eq:Morse_system} and find $a,t,V_e$ for some given real finite number $V,g,H$ in the limit  $D_e\rightarrow\infty $.

For $V,g,H$ to be finite as $D_e$ tends to infinity, we obtain from the first equation that $(1-t)$ should be infinitesimal, and from the third equation that $a$ should also be infinitesimal.

In the limit of $a \rightarrow 0$ and $ (1-t) \rightarrow 0$ we can make the following approximations:
\begin{equation}
\begin{aligned} 
(1-t)= (1-e^{-a(r-r_e)}) \approx a(r-r_e)  \\ 
t\approx 1  \\ 
(2t-1) \approx 1
\end{aligned}
\end{equation}

With these approximations the system take form:
\begin{equation}
\begin{cases} 
V\approx D_e a^2(r-r_e)^2 +V_e \\ \\
g\approx 2D_e a^2(r-r_e)  \\ \\
 H\approx 2D_e a^2
\end{cases} 
\end{equation}
Which are the equation of a parabola with a second order coefficient $k=D_ea^2=H/2$ and its first and second derivatives. Therefore in the limit $D_e\rightarrow\infty$ the Morse and the harmonic interpolation give identical results.

\subsection{Dependence on $D_e$ and comparison with the Newton-Raphson method}
In order to interpolate a molecular energy surface (knowing $V,g,H$ in some point) to a Morse potential we need also to know the value of the bond dissociation energy ($D_e$) or an approximation of it.
For diatomics $D_e$ can be calculated subtracting the energy in the starting point $E(R_0)$ from sum of the energy of the two isolated atoms. 

\begin{figure}
    \centering
    \includegraphics[width=\linewidth]{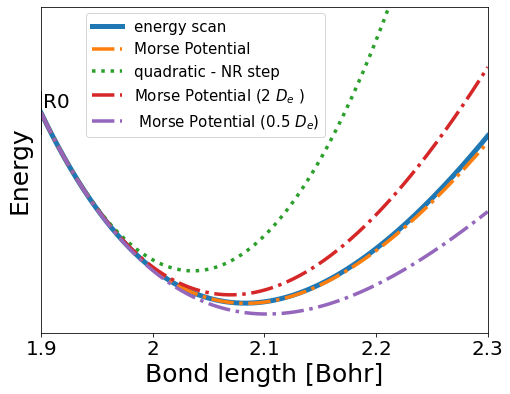}
    \caption{For the CO molecule at the starting interatomic distance $R_0=1.9$ Bohr were calculated energy, gradient and hessian. With that was possible to take an optimization step using the NR method, and approximating $D_e \approx E(r=\infty)-E(r=R_0)$  the energy was interpolated to a Morse potential. The plot also shows the Morse potential interpolated using also a dissociation energy two times bigger and a dissociation energy equal to the half of $D_e$ in order to show the effect of this parameter on the shape of the dissociation curve.}
    \label{fig:Morse_CO_1}
\end{figure}

In Figure \ref{fig:Morse_CO_1} we compare the results of the Morse interpolation obtained using the calculated $D_e$, with the one obtained choosing $D_e$ to be twice and one half the calculated value, and with the harmonic approximation.
A lower dissociation energy leads to a wider energy hole, a higher dissociation energy makes the hole narrower and in the limit of $D_e\rightarrow\infty$ the curve tends to the quadratic approximation. 
All the interpolated Morse potentials approximate the true energy curve better than the parabola, if we manage to get an empirical overestimation of the dissociation energy we are sure to get more accurate results using Morse interpolation than using harmonic approximations.

\subsection{Transformation in redundant internal coordinates}
The transformation of  the coordinates and energy  derivatives from Cartesian to  redundant internal coordinates, was performed using the PyBerny libraries.\cite{Pyberny}
Internal coordinates were chosen according to criteria outlined in reference \cite{Peng}:
\begin{itemize}
    \item All bonds shorter than 1.3 times the sum of covalent radii are created 
    \item All angles greater than 45° are created.
    \item All dihedrals with 1–2–3, 2–3–4 angles both greater than 45° are created. If one of the angles is zero, so that three atoms lie on a line, they are used as a new base for a dihedral. This process is recursively repeated 
\end{itemize}

The expression of energy derivatives (gradient and Hessian) are the one proposed by Pulay and 
Fogarasi \cite{fogarasi1992calculation,pulay1992geometry}.
Being $B$ the Wilson matrix 
\begin{equation}
    B_{ij} :=\frac{\partial q_i}{\partial x_j} 
\end{equation}
The transformation of the gradient from internal coordinates ($q$) to Cartesian ones ($x$) is made:
\begin{equation}
    \mathbf{g}_x=B^T \mathbf{g}_q
\end{equation} 
to invert the relationship we use the pseudoinverse of $B$:

\begin{align}
    B_{inv} = B^T(B^T B)^{-1} \\
    \mathbf{g}_q=B_{inv}^T \mathbf{g}_x 
\end{align}

While for the Hessian if $B'_{ijk}=\partial q_i /\partial x_j \partial x_k$  than the following transformations can be made :

\begin{equation}
    H_x=B^T H_q B+B'^T\mathbf{g}_q
\end{equation} 
and the other way around:
\begin{equation}
    H_q=B_{inv}^T (H_x-B'^T\mathbf{g}_q)B_{inv}    
\end{equation}


\subsection{Application of the Morse potential interpolation to non linear molecules}

Using internal coordinates ($\mathbf{q}$), gradient $\mathbf{g}=\mathbf{g}_q$ and Hessian $H=H_q$ also expressed in internal coordinates.
The Newton Raphson step $\Delta \mathbf{q}^{NR}$ is calculated solving:
\begin{equation}
\label{eq:SI_NR_step}
    H \Delta \mathbf{q}^{NR}=-\mathbf{g} 
\end{equation} 

The interpolation to a Morse potential can be included as a correction $\Delta \mathbf{q}^{Corr.}$ to the newton Raphson step:
\begin{equation}
    \Delta \mathbf{q} =\Delta \mathbf{q}^{NR} + \Delta \mathbf{q}^{Corr.}
\end{equation}  

For every bond $b$ we can insert $\mathbf{g}_b$ and $H_{bb}$ in Eqs. \ref{Eq:Morse_system}, we obtain then the Morse rearrangement $\Delta \mathbf{q}_b^{Morse} $ and the Morse relaxation energy $\Delta E^{Morse}$.

To obtain the correction for bonds we subtract from $\Delta \mathbf{q}^{Morse}$ the Newton Raphson step for $b$ as an independent coordinate.

\begin{equation}
    \Delta \mathbf{q}_b^{Corr.} = \Delta \mathbf{q}_b^{Morse} -(-\mathbf{g}_b/H_{bb}) \ \ \  \ b \ is \ a \ bond. 
\end{equation}

A change in bond lengths affects the rearrangement of angles and dihedrals through the non diagonal part of the Hessian, for any angle or dihedral $a$ Eq. \ref{eq:SI_NR_step} should hold also after the correction: 
\begin{equation}
\begin{aligned}
    H (\Delta \mathbf{q}^{NR}+\Delta \mathbf{q}^{Corr.}) &=-\mathbf{g} \\
    \sum_i^{angles,\atop bonds}H_{a'i} \Delta \mathbf{q}^{Corr.}_i&=0 \\
    \sum_{a}^{angles}H_{a'a} \Delta \mathbf{q}^{Corr.}_{a}&=-\sum_{b}^{bonds}H_{a'b} \Delta \mathbf{q}^{Corr.}_{b} \\
    \Delta \mathbf{q}_a^{Corr.} &=-H_{aa'}^{(-1)} H_{a'b}\Delta\mathbf{q}_b^{Corr.}
\end{aligned}
\label{eq:Morse_as_correction2_si} 
\end{equation}
The Newton Raphson relaxation energy can be expressed as:
\begin{equation}
    \Delta E^{NR}=-\frac{1}{2}\Delta\mathbf{q}^TH\Delta\mathbf{q}
\end{equation}

The energy can be corrected if for every bond we substitute the Newton Raphson energy $-H_{bb}(\Delta\mathbf{q}_b^{Morse})^2$ with the energy from the Morse potential interpolation $\Delta E^{Morse}_b$ :
\begin{equation}
    \Delta E=-\Delta E^{NR} +\sum_b^{bonds}(-(-\frac{1}{2} H_{bb}(\Delta\mathbf{q}_b^{Morse})^2) + \Delta E^{Morse}_b)
\end{equation}

\section{References}

\bibliography{bib_files/BasisRef,bib_files/Alchemy_vLg,bib_files/Alchemy_Others,bib_files/software,bib_files/Other_works,bib_files/QML,bib_files/derivatives,bib_files/Geomopt}